\documentclass[a4paper]{iopart}
\usepackage[cp850]{inputenc}
\usepackage{mathptmx}
\usepackage{geometry}
\usepackage{graphicx}
\usepackage{multicol}
\usepackage[
   colorlinks,        
   linkcolor=blue,   
   filecolor=blue,   
   citecolor=blue   
]{hyperref}
\usepackage{textcomp}
\begin{document}

\title[Electrical-transport characteristics of La$_{0.7}$Ce$_{0.3}$MnO$_3$ films]{Electrical-transport characteristics of as-grown and oxygen-reduced 
La$_{0.7}$Ce$_{0.3}$MnO$_3$ films: 
calculation of hopping energies, Mn valences, and carrier localization lengths}

\author{A Thiessen$^1$, E Beyreuther$^1$,
R Werner$^2$, R Kleiner$^2$, D~Koelle$^2$ and L~M~Eng$^1$}

\address{$^1$Institut f\"ur Angewandte Photophysik, 
Technische Universit\"at Dresden, D-01062 Dresden, Germany \\
$^2$Physikalisches Institut and Center for Collective 
Quantum Phenomena in LISA$^+$,
Universit\"at T\"ubingen, Auf der Morgenstelle 14, D-72076 T\"ubingen, Germany}

\eads{\mailto{andreas.thiessen@iapp.de}, \mailto{elke.beyreuther@iapp.de}}

\begin{abstract}
Presently, cerium-doped LaMnO$_3$ is vividly discussed as an  electron-doped 
counterpart prototype to the well-established hole-doped mixed-valence manganites. 
Here, La$_{0.7}$Ce$_{0.3}$MnO$_3$ thin films of different thicknesses, degrees 
of CeO$_2$ phase segregation, and oxygen deficiency, grown on SrTiO$_3$ single 
crystal substrates, are compared with respect to their 
resistance-vs.-temperature (R vs. T)
behavior from 300~K down to 90~K. While the variation of the film thickness (and thus the 
degree of epitaxial strain) in the range between 10~nm and 100~nm has only 
a weak impact on the electrical transport, the degree of oxygen deficiency 
as well as the existence of CeO$_2$ clusters can completely change the 
type of hopping mechanism. This is shown by fitting the respective 
\textit{R-T} curves with three different transport models (adiabatic polaron hopping, 
Mott variable-range hopping, Efros-Shklovskii variable-range hopping), 
which are commonly used for the mixed-valence manganites. 
Several characteristic transport parameters, such as the hopping 
energies, the carrier localization lengths, as well as the Mn valences 
are derived from the fitting 
procedures.
\end{abstract}



\section{Introduction}
\label{sec_Introduction}

Doped rare-earth manganites (also referred to as 
mixed-valence manganites \cite{coey_mixed-valence_1999} or 
colossal magnetoresistive (CMR) manganites 
\cite{haghiri-gosnet_cmr_2003}) 
have been in the focus of intense scientific 
interest for several decades due to their peculiar magnetic/electrical 
phase diagrams, their magnetoresistance, and the subtle interplay between 
different (lattice, charge, spin, orbital) degrees of freedom. In 
the early years, fundamental issues such as the explanation of 
the relation between electrical transport and magnetic order in bulk 
crystals were in the focus
\cite{volger_further_1954,anderson_considerations_1955,millis_fermi-liquid--polaron_1996}. 
Later, especially 
with the decisive progress in thin film preparation 
\cite{helmholt_giant_1993}, a number of further intriguing issues were 
under debate, e.g., questions 
related to the functionalization of heterostructures and interfaces, 
electronic phase separation \cite{dagotto_colossal_2001}, 
or the response to light excitation \cite{beyreuther_large_2009}.

Concerning the research field of functional heterostructures, 
such as all-manganite junctions 
\cite{tanaka_giant_2001,mitra_p-n_2001,mitra_observation_2003}, 
the possibilities of preparing 
electron-doped counterparts to the well-known hole-doped (lanthanum) 
manganites have been investigated for almost two decades. 
In the well-established case of the hole-doped La manganites, \emph{divalent} cations 
such as Ca$^{2+}$ or Sr$^{2+}$ 
replace part of the La$^{3+}$ ions, and concurrently force an equivalent number of Mn ions 
into the 4+ state 
to preserve charge neutrality. The resulting mixed 3+/4+ Mn valence goes 
along with the double exchange mechanism connected to a number 
of manganite-typical properties such as a metal-insulator transition associated 
with a ferromagnetic-paramagnetic phase transition and (at least in thin films) 
the colossal magnetoresistance
effect. On the other hand, substituting part of the La$^{3+}$ ions in the 
LaMnO$_3$ host lattice by \emph{tetravalent} cations such as Ce$^{4+}$, 
Te$^{4+}$, or Sb$^{4+}$ leads to a nominal electron-doping, connected with a 
mixed Mn$^{2+}$/Mn$^{3+}$ valence, the latter being supposed to be equivalent 
to the mixed 3+/4+ case with respect to the double-exchange 
mechanism \cite{Schlottmann2008}. Here, 
the cerium-doped LaMnO$_3$, typically in the form of epitaxial 
La$_{0.7}$Ce$_{0.3}$MnO$_3$ (LCeMO) thin films 
\cite{raychaudhuri_phase_2003}, plays the role of
a model compound for studying electron-doped lanthanum 
manganites. However, as shown by a growing body of experimental data, as-prepared 
LCeMO films tend to oxygen excess and thus an effective hole-doping (see 
e.g.~\cite{bey06} and refs. therein), while 
oxygen reduction by post-deposition annealing can indeed produce 
Mn valences below 3$+$, typically followed by the loss of the 
manganite-typical metal-insulator transition, producing films which 
are insulating across the whole temperature range \cite{beyreuther_large_2009}. 
A recent study by Middey et al.~\cite{Middey2014} demonstrated that LCeMO thin films grown layer-by-layer with pulsed laser interval deposition showed already in the as-prepared state electron doping but remained insulating in the whole temperature range, too.
Attributed to the large variety of chemical compositions of manganite 
compounds (of which a number still waits for an in-depth analysis 
of its physical parameters) 
comprehensive electrical-transport data of this electron-doped state in 
La$_{0.7}$Ce$_{0.3}$MnO$_3$ films has rarely been reported to date, which
is exactly the starting point of the work presented here.

\section{Experimental}

Pulsed laser deposition (PLD) was employed to grow three different 
La$_{0.7}$Ce$_{0.3}$MnO$_3$ films on SrTiO$_3$ 
(100) single crystal substrates. 
The exact growth conditions and the results of the structural analysis 
can be found in reference~\cite{wer09}, the most important 
parameters (sample label / film thickness / oxygen partial 
pressure during PLD / result of the x-ray diffraction analysis) are listed here: 

\begin{itemize}
\item B / 30 nm / 0.25 mbar / single-phase,\footnote{Though appearing single-phasic 
in the XRD analysis, a transmission electron microscopy study showed 
the existence of nanoscopic CeO$_2$ clusters for samples equivalent to B and C 
\cite{wer09}.}
\item C / 100 nm / 0.25 mbar / single-phase,
\item D / 100 nm / 0.03 mbar / CeO$_2$ clusters.
\end{itemize}

As in our previous works on LCeMO 
\cite{bey06,thiessen_mn2+/mn3+_2014}, which focused on the investigation of the 
Mn valence by photo\-emission spectroscopy, the 
samples were investigated in different states of oxygen content: 

\begin{itemize}
\item the \emph{as-prepared} state,
\item a \emph{slightly reduced} state by 
      heating in ultrahigh vacuum (10$^{-9}$~mbar) at 480$\symbol{23}$C for 1~h, and
\item a \emph{highly reduced} state by 
      heating in ultrahigh vacuum (10$^{-9}$~mbar) at 700$\symbol{23}$C for 2~h. 
\end{itemize}

To link the present measurements to our previous investigations, see 
refs.~\cite{beyreuther_large_2009,bey06,thiessen_mn2+/mn3+_2014}, 
the transport measurements 
of the as-prepared samples were compared to the results of a 10-nm LCeMO thin film, referred to as 
\emph{sample A}, 
which was already subject of these former studies.

The resistance $R$ of the LCeMO films was measured in the dark as a function of 
the temperature $T$ in a 
nitrogen-cooled cryostat (Optistat DN by Oxford Instruments). Electrical 
contacts with a 3-mm distance were made by conductive silver paste on the 
LCeMO surfaces. To prevent 
leakage currents but guarantee good thermal contact 
the samples were mounted by conductive silver paste on a 
sapphire plate, while the 
temperature was measured by a Pt-100 resistor mounted on the same sapphire 
plate directly besides the sample.
Since in all cases the LCeMO
resistance was orders of magnitude higher than the wiring resistance, the
measurements were performed in two-point geometry using 
a Keithley 6517B electrometer with a constant measuring voltage of 1~V, unless 
stated otherwise. 

\section{Theoretical background}

For convenience, and also since the relevant information appears to be quite 
scattered throughout the literature, we summarize the most important
knowledge concerning charge transport models 
for the insulating paramagnetic phase of mixed-valence manganites.

In the insulating phase the charge carriers are -- at least partially 
-- localized, and thus are not subject to a coherent band transport as it is 
the case in 
conventional semiconductors, but \emph{move diffusively by thermally activated 
hopping between the Mn ions}. Thus the temperature dependence of the 
conductivity $\sigma(T)$ is inherently related to the hopping frequency 
$p(T)$ via:

\begin{equation}
	\sigma(T)=\frac{x(x-1)}{c^3}\frac{e^2r^2}{k_B T} p(T) \quad ,
\end{equation}

with $c$ being the lattice constant, $x$ the Mn$^{4+}$ or Mn$^{2+}$ concentration for hole or electron doped compounds, respectively, $e$ the 
elementary charge, $k_B$ the Boltzmann constant, 
and $r$ the distance between two states assumed to be equivalent 
within $k_B T$ \cite{wor98}. 
For the temperature dependence of the 
hopping probability (hopping frequency) $p(T)$ several 
variable-range hopping (VRH) \cite{coe95,vir97,wag98} models, 
adiabatic polaron hopping (APH) \cite{wor98,sal01,edw02} models, 
or combinations of both \cite{sun00,lai02} were reported in the literature. 
Briefly, the concepts behind APH and VRH are explained in the following:

\subsection{Adiabatic polaron hopping}

Due to strong electron-phonon coupling, the carriers distort the 
crystal lattice in their 
surroundings. Whenever the characteristic energy of this interaction is 
low compared to the conduction-band width, the carriers can further contribute to the 
coherent band transport, but the induced lattice polarization will follow 
the carriers, the latter thus exhibiting a large effective mass. Both, the 
electron and the lattice polarization form a quasiparticle, the so-called 
\emph{large polaron} \cite{frohlich_properties_1950}. For mixed-valence 
manganites, however, the electron-phonon coupling is strong compared 
to the bandwidth. Thus, the carriers become trapped in potential wells 
built by the lattice distortions, hence revealing \emph{small polarons}. 
In this case, thermally activated hopping may occur. Since lattice movements 
are slow compared to the hopping velocities, the the term 
\emph{adiabatic} is used for this process. 

Within the framework of this APH model, 
which has been described by Edwards 
\cite{edw02} in detail, the following relation is derived for $\sigma$:

\begin{equation}
	\sigma(T)\propto T^{-1}
	\exp\left(-\frac{E_h}{k_B T}\right)   \quad  ,
\end{equation}

with $E_h$ being the hopping energy, connected to the polaron 
binding energy $E_p$ via $E_h=E_p/2$ for oxides in general 
\cite{tsuda_electronic_conduction_2000} and via $E_h=E_p/4$ 
for manganites \cite{edw02}. 

\subsection{Variable-range hopping}

If the Fermi level is located within 
a part of the conduction band which is localized by disorder,
charge transport is possible either by thermal excitation 
into an empty and non-localized part of the band or by thermally activated 
hopping between localized states. The first case is relevant at high 
temperatures and results in the common thermally-activated band transport, 
while the latter case leads to the concept of the variable-range hopping as is 
relevant for the manganites. The term \emph{variable-range} is derived from 
the fact that in this scenario the hopping distances (either between 
next neighbours or larger) can be, from energetic considerations, 
different, depending on the temperature.

The VRH theory, as described by Mott in ref.~\cite{mot12}, 
results in the following 
temperature-dependent conductivity:

\begin{equation}
\label{sigma_VRH_Mott}
	\sigma(T)\propto T^{-1/4}
	\exp\left[-\left(\frac{T_M}{T}\right)^{1/4}\right] \quad  ,
\end{equation}

with the Mott temperature $T_M$ calculated as:

\begin{equation}
\label{T_M}
	T_M=\frac{\alpha_M}{k_B N(E_F) a^3}        \quad .
\end{equation}

Here, $\alpha_M$ is a variable number between 10 and 27 (depending on the 
exact derivation \cite{sea74}), $a$ the localization 
length of the charge carrier, and $N(E_F)$ the 
density of states 
at the Fermi level. $T_M$ is typically extracted as a 
fit parameter from the experimental conductivity-vs.-temperature data. From 
$T_M$ the corresponding hopping energy $E_M$ and the hopping distance $r$ 
can be obtained according to:

\begin{equation}
\label{E_M}
	E_M=k_B T \left(\frac{T_M}{T}\right)^{1/4} \quad ; \quad
	r=a \left(\frac{T_M}{T}\right)^{1/4} \quad .
\end{equation}

The Mott-VRH model was modified by Efros and Shklovskii by including the 
influence of electron correlation on the density of states which leads to the formation of a parabolic gap  in the densisty of states at the Fermi energy \cite{efr75}, the so called Coulomb gap $\Delta_C$. The 
temperature dependence of $\sigma$ then reads as:

\begin{equation}
	\sigma(T)\propto T^{-1/2}
	\exp\left[-\left(\frac{T_{ES}}{T}\right)^{1/2}\right] \quad , 
\end{equation}

with $T_{ES}$ being the Efros-Shklovskii temperature:

\begin{equation}
\label{T_ES}
	T_{ES}=\frac{2.8 e^2}{k_B \epsilon_0 \epsilon_r a}     \quad .
\end{equation}

Unlike $T_M$, $T_{ES}$ is independent of the density of states but depends on 
the dielectric constant $\epsilon_r$. In the Efros-Shklovskii (ES) VRH model, 
the hopping energy and distance are calculated as follows:

\begin{equation}
\label{E_ES}
	E_{ES}=k_B T \left(\frac{T_{ES}}{T}\right)^{1/2} \quad ; \quad
	r=a \left(\frac{T_{ES}}{T}\right)^{1/2} \quad .
\end{equation}

The ES-VRH model becomes valid as soon as the hopping energy $E_M$ is smaller 
than the Coulomb gap $\Delta_C$, which is defined as:

\begin{equation}
\label{Delta_C}
	\Delta_C = \frac{e^3 N(E_F)^{1/2}}{(\epsilon_0 \epsilon_r)^{3/2}}\quad .
\end{equation}


For the interpretation of the hopping transport within a 
very large temperature range in CMR manganites,
 Laiho et al. \cite{lai02, Laiho05} have extended
the ES-VRH model by taking into 
account some peculiarities like 
the high temperatures at which VRH is observed in CMR manganites.
Additionally, they took the 
influence of a temperature dependent rigid polaronic 
pseudo gap in the densitiy of states \cite{edw02,biswas_density_1999}, $\Delta_{pg}$, into account
 and deduced the following 
temperature dependence of the conductivity:

\begin{equation}
\label{sigma_mod_ES_VRH}
	\sigma(T) \propto T^{-9/2}
	\exp\left[-\left(\frac{T_0}{T}\right)^{1/2}\right] \quad , 
\end{equation}

with $T_0$ being calculated as:

\begin{equation}
\label{T_0_mod_ES_VRH}
	T_0=\left[\frac{\Delta_{pg}}{2k_B T^{1/2}}+
	\left(\frac{\Delta_{pg}^2}{4k_B^2 T}+T_{ES} 
	  \right)^{1/2}\right]^2\quad . 
\end{equation}

\subsection{Experimental findings for the manganites}

Reviewing a number of currently available experimental studies of the 
temperature dependent transport in the insulating phase,
the data was interpreted in terms of the APH model in 
refs.~\cite{wor98,jai96a,sny96,jak98,ter98}, while 
other authors \cite{vir97,wag98,ter96,jai96b,ter97} employed the 
Mott-VRH model or the ES-VRH model \cite{vir97,wag98,lai02}. 

In general one can summarize that (i) for temperatures 
above 300~K the transport is determined by the carrier-lattice 
interaction, i.e., the thermally activated diffusion of small 
polarons being described by the APH model, while (ii) for lower temperatures 
carrier localization by disorder leads to hopping transport, which 
is described by the several variable-range hopping models. 

\begin{figure*}
\centering
\includegraphics[width=0.9\textwidth]{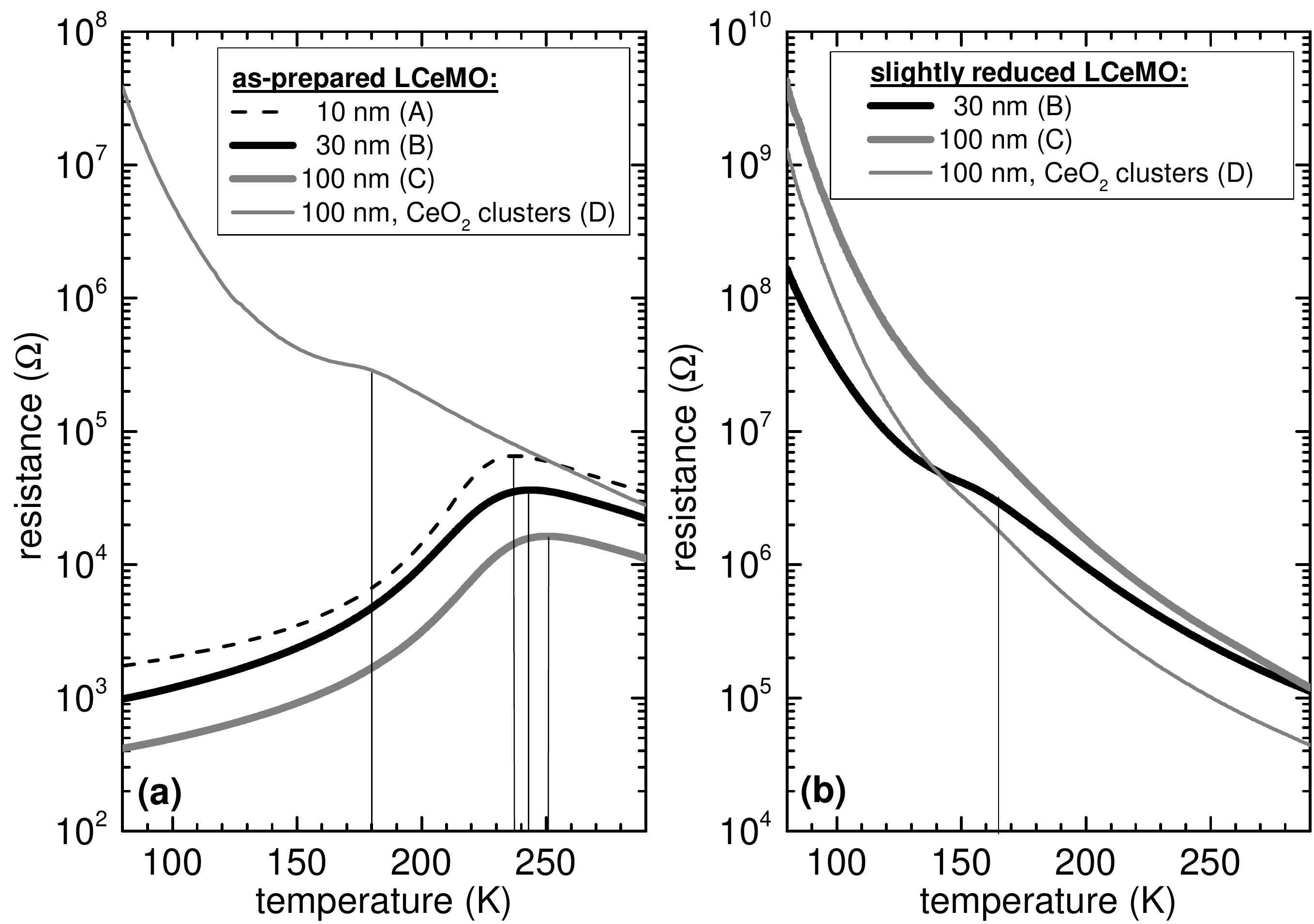}
\caption{\label{fig1}Resistance-vs.-temperature characteristics of 
the \emph{as-prepared} LCeMO films (a), and the \emph{slightly reduced} 
films (b). The vertical lines indicate the positions of the 
MITs, $T_{MIT}$, or the beginning of a plateau, $T_{pl}$, respectively.}
\end{figure*}

\section{Results}

First, the resistance-vs.-temperature curves of the 
\emph{as-prepared} films were recorded, as shown in figure~\ref{fig1}(a). 
The samples A\footnote{The \textit{R-T} curve of sample A is shown for the 
as-prepared case only 
to make the link to our previous investigations 
\cite{bey06,beyreuther_large_2009}, where this 
10-nm film was in the focus.}, 
B, and C show metal-insulator transitions (MITs)
at 237~K, 243~K, and 251~K, respectively.  The MIT temperatures ($T_{MIT}$) were 
extracted by smoothing 
and differentiating the \textit{R-T} curves, looking for the points where the 
first derivatives of the respective \textit{R-T} curve become zero. The values obtained
are typical for LCeMO films with a hole doping by oxygen excess. The 
variation of the transition temperature with the film thickness is a well-known 
phenomenon and can be attributed to stronger relaxation of substrate-induced strain with growing film thickness.

Solely sample D does not exhibit an MIT, but a weak plateau at 180~K, which is 
commonly interpreted as a transition between a paramagnetic-insulating to a 
ferromagnetic-insulating phase and points towards an insufficient doping 
\cite{seeger_charge_1999}. Obviously, the low oxygen partial pressure during 
the deposition of film D is the origin for the deviating \textit{R-T} characteristics. 
As proven previously on an equivalent film \cite{wer09}, 
the MIT can be recovered by appropriate annealing under oxygen atmosphere.

The \textit{R-T} curves in the insulating range between 300~K and 270~K can be reproduced 
by both the adiabatic polaron hopping (APH) and the variable-range hopping (VRH) 
model. 
A regression analysis assuming the APH model reveals polaron hopping energies 
of 115~mV for samples A, B, C, and 144~mV for sample D. According to the 
empirical relation, originally derived for La$_{0.7}$Ca$_{0.3}$MnO$_{3-\delta}$  
\cite{ter98}:

\begin{equation}
\label{hopping_energy_vs_Mn_valence}
E_h=c(Mn^{3+})\cdot 627~\mbox{meV} - 377~\mbox{meV} \quad ,
\end{equation} 

between the Mn$^{3+}$ concentration $c(Mn^{3+})$ and the activation energy 
$E_h$ for the APH 
process, we obtain a Mn$^{3+}$ concentration of 78\% for samples 
A, B, C. Since the \emph{as-prepared} films had been shown to suffer from overoxygenation and concomitant hole doping \cite{bey06, wer09}, it is reasonable to assume that the residual 22\% of Mn ions have a valence of 4+. The corresponding average Mn valence is 3.22, being dramatically larger 
than the nominal Mn valence of 2.7, but being in good accordance with 
former results obtained on 10-nm thick as-prepared LCeMO films 
\cite{bey06}. For sample D we obtain 
83\% Mn$^{3+}$ and -- due to the lower oxygen partial pressure 
during the film deposition -- a smaller Mn valence of 3.17, which is 
still far above the nominal value.

The resistances of the \emph{slightly reduced} films are 
decisively higher than in the as-prepared case 
[figure~\ref{fig1}(b)]. Moreover, there are no 
clear MITs visible any longer, but a small plateau beginning at 165~K 
for sample B, while for C and D plateaus are hardly visible. 
Between 300~K and the plateau temperature the APH model gives 
hopping energies of 140~meV and 165~meV for B and C, respectively, which are 
higher than in the as-prepared cases and correspond to Mn valences 
of 3.18 (sample B) and 3.14 (sample C). These values are in good agreement with 
those derived from previous XPS results \cite{thiessen_mn2+/mn3+_2014}. 
For sample D the same hopping energy as in the as-prepared case (144~meV) is 
extracted from the APH model, while the Mn valence from the present
transport data (3.17) is lower than the value from XPS (3.31). One may speculate that 
the surface sensitivity of the XPS analysis combined with the higher disorder 
of sample D and the possible inhomogeneous oxygen diffusion during the 
oxygen-reducing annealing process are the origin for this discrepancy.

For clarity, all numeric results for the as-prepared and slightly reduced films are 
summarized in table~\ref{tab_1}.

\begin{table*}
\caption{\label{tab_1}\emph{As-prepared} and \emph{slightly reduced} 
LCeMO films: Overview of the characteristic temperatures 
(either the temperature of the metal-insulator transition $T_{MIT}$ or 
the plateau temperature $T_{pl}$), 
the polaron hopping energies $E_h$ extracted from fitting with the APH model, and 
the corresponding Mn valences $V_{Mn}$ according to 
eq.~(\ref{hopping_energy_vs_Mn_valence}). For the slightly reduced 
films the Mn valences from our former XPS analysis \cite{thiessen_mn2+/mn3+_2014} 
are given in brackets for comparison.}
\renewcommand{\arraystretch}{1.3}
\begin{indented}
\lineup
\item[]\begin{tabular}{@{}crcccc|ccc}
\br
 & &	&\multicolumn{3}{c}{as-prepared state} &\multicolumn{3}{c}{slightly reduced state} \\
 \cline{4-6}\cline{7-9}
sample & thickness & CeO$_2$ clusters & $T_{MIT}, T_{pl}$ & $E_h$& $V_{Mn} $ & 
$T_{pl}$ & $E_h$& $V_{Mn} (V_{Mn}^{XPS})$ \\
 & (nm) &		& (K)		& (meV)		&	& (K)	& (meV) & \\
\hline
B & 30 & nanoscopic & 243 (MIT) & 115 & 3.22 & 165 & 140 & 3.18 (3.19) \\
C & 100 & nanoscopic & 251 (MIT) & 115 & 3.22 & -- & 165 & 3.14 (3.10) \\
D & 100 & microscopic & 180 (pl.) & 144 & 3.17 & -- & 144 & 3.17 (3.31)\\
\br
\end{tabular}
\end{indented}
\end{table*}

\begin{table*}
\caption{\label{tab_2}\emph{Highly reduced} LCeMO films: Overview of the fitting parameters 
derived from the APH model (columns 2--4) in the higher temperature range and from the 
respective (most suitable) VRH model in the lower temperature range (columns 5--10).}
\renewcommand{\arraystretch}{1.3}
\begin{indented}
\lineup
\item[]\begin{tabular}{@{}lccr|ccccc}
\br
&\multicolumn{3}{c}{adiabatic polaron hopping} &\multicolumn{5}{c}{variable-range hopping} \\
 \cline{2-4}\cline{5-9}
sample & range & $E_h$ & $V_{Mn} (V_{Mn}^{XPS})$ & type & range& $T_{ES}, T_M$ & $a$&  $\Delta_C$ \\
 & (K) & (meV) & & 	& (K)	& (K)	& (nm) & (eV) \\
\hline
B & 300..210 & 281 & -- (2.81) & ES & below 210 & 1.06$\times 10^5$ & 0.3-0.6 & 0.41 \\
C & 300..210 & 228 & 2.96 (2.81) & ES & below 210 & 1.22$\times 10^5$ & 0.3-0.5 & 0.44 \\
D (1 V) & 300..260 & 52 & 2.68 (2.83) & Mott & 260...80 & 6.61$\times$10$^8$ & 0.09 & -- \\
D (5 V) & 300..280 & 62 & 2.70 (2.83) & Mott & 260...190 & 3.54$\times$10$^8$ & 0.11 & -- \\
\br
\end{tabular}
\end{indented}
\end{table*}

\begin{figure*}
\centering
\includegraphics[width=0.85\textwidth]{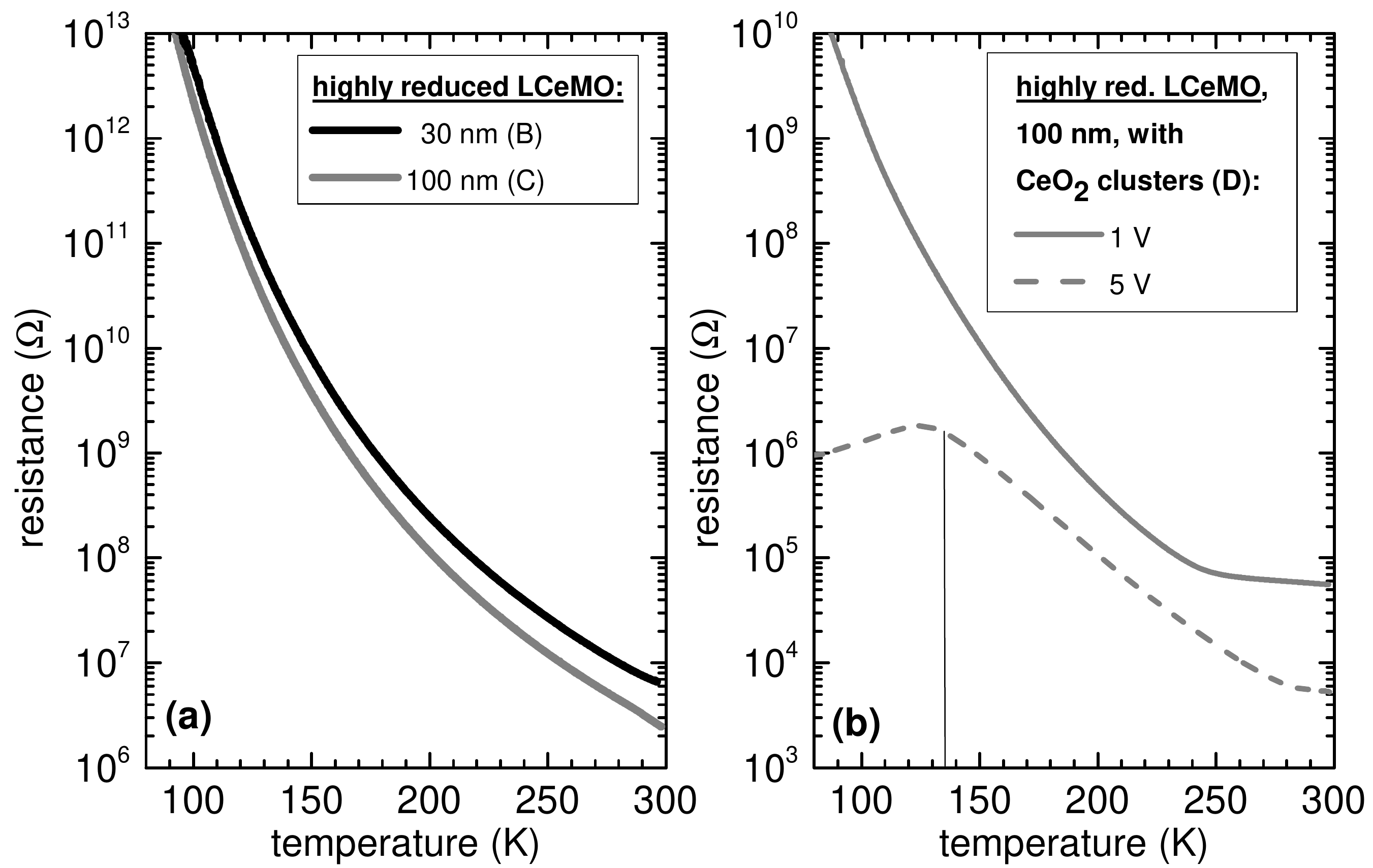}
\caption{\label{fig2}Resistance-vs.-temperature characteristics of 
the \emph{highly reduced} LCeMO films: (a) Samples B and C; (b) sample 
D at two different measuring voltages (1~V, 5~V), showing 
an unexpected, pronounced electroresistive effect.}
\end{figure*}

As expected, the resistances of the \emph{highly reduced} LCeMO films 
tend to be higher than in the slightly-reduced samples, 
see figures~\ref{fig2}(a),(b). As an exception, the resistance of sample D at 1 V 
is higher in the slightly-reduced case between 275~K and 200~K, but within the same order 
of magnitude. At a measuring voltage of 1~V, all samples exhibit 
an insulating behavior in the whole investigated temperature range and no sign 
of a MIT. 

Interestingly -- and indeed electroresistive effects in LCeMO were reported in the past 
\cite{bajaj_electroresistive_2007} -- the resistance of film D strongly depends on the 
measuring voltage. As visible in figure~\ref{fig2}(b), in the whole 
temperature range investigated here, the resistance at 5~V is 
clearly lower than at 1~V. Furthermore, at 5~V, an MIT 
appears at 135~K. Being beyond the scope of the present work, 
the clarification of the origin of this 
electrical-field induced MIT (intrinsic to 
the film or substrate-related) by comparative 
magnetoresistance investigations is currently in progress 
\cite{thiessen_conductivity_2014}. Note that the electroresistive effect 
turned out to be non-persistent and was observed solely in sample D.

\begin{figure*}
\centering
\includegraphics[width=0.9\textwidth]{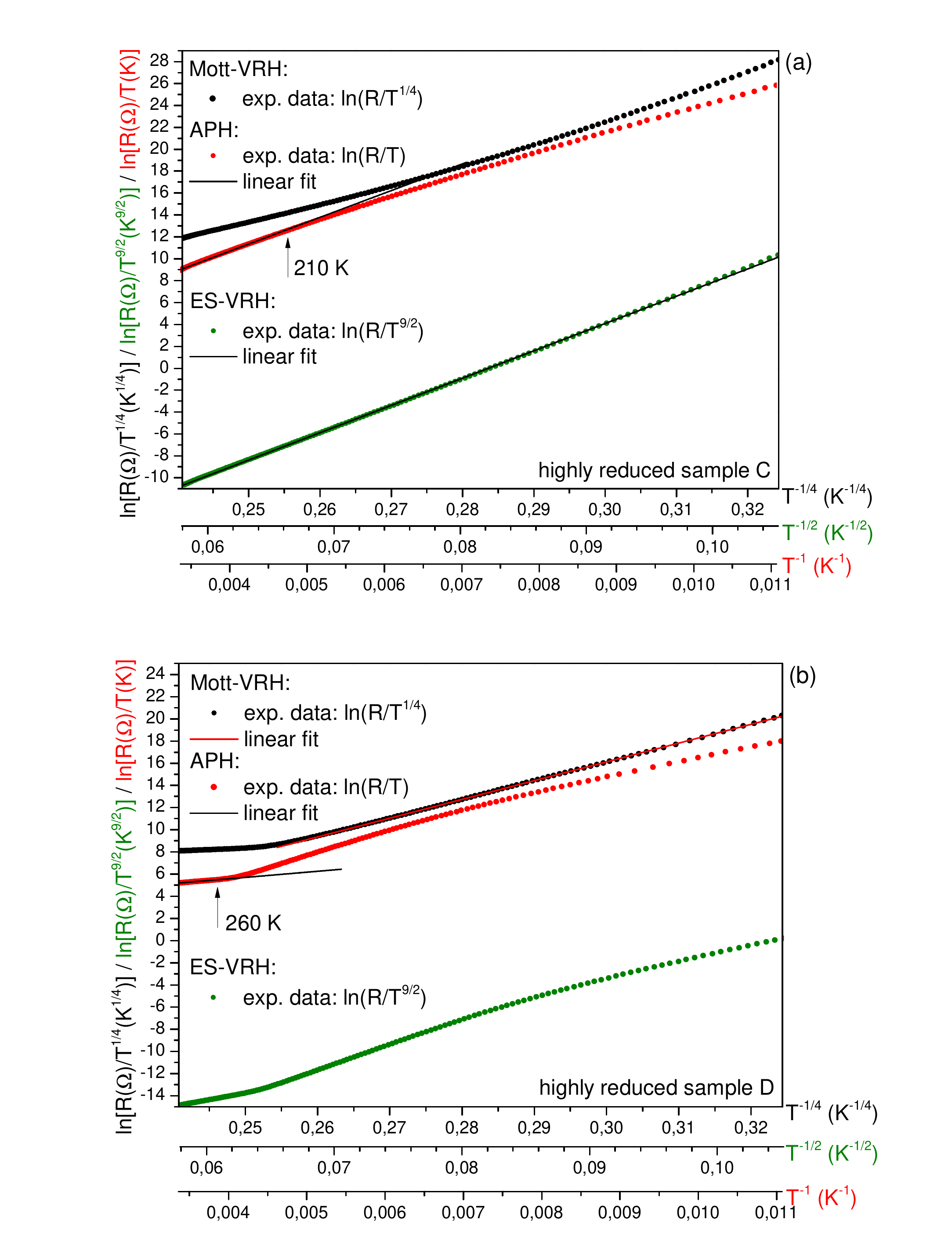}
\caption{\label{fig3} \emph{Highly reduced} LCeMO films: 
Plots of the logarithms $\ln(R/T)$, 
$\ln(R/T^{1/4})$, and $\ln(R/T^{9/2})$ against $T^{-1}$, $T^{-1/4}$, and 
$T^{-1/2}$, respectively, for T between 300~K and 90~K in order to check the validity of either the 
APH, the Mott-VRH, or the modified ES-VRH model -- exemplarily shown for samples 
C (a) and D (b); the data shown here was recorded at 1~V for both samples. 
The vertical arrows indicate the temperature below which one of the 
VRH models had to be used.
For the exact fitting results refer to table~\ref{tab_2}.}
\end{figure*}

Let us now, as in the cases of the as-prepared and the slightly-reduced films, 
continue with a detailed analysis of the resistances by curve fitting. 
For samples B and C, the \textit{R-T} curves can be modeled by the APH model 
in the range between 300~K and 210~K. From this model we obtain polaron hopping 
energies of 281~meV (B) and 228~meV (C). 

 According to 
eq.~(\ref{hopping_energy_vs_Mn_valence}) 
 these hopping energies would be equivalent 
to a Mn$^{3+}$ concentration of 96\% for C, which is clearly different from the 81\% 
from the former XPS investigation \cite{thiessen_mn2+/mn3+_2014}, while 
for B we get an unphysically high 
Mn$^{3+}$ concentration of 105\%, which causes some doubts on the validity 
of eq.~(\ref{hopping_energy_vs_Mn_valence}), at least for the case of the highly reduced 
LCeMO films.
Concerning the discrepancy of the XPS- and the transport related 
Mn$^{3+}$ concentrations 
of sample C one could argue that there is a gradient of the Mn$^{3+}$ 
concentration from the film surface towards deeper regions of the film 
and that XPS as a 
surface-sensitive method naturally detects only the value correct for 
the first few nanometers beneath the surface. In other words, the 
films are stronger oxygen-deficient at the surface, which is indeed 
in agreement with the former x-ray absorption (XAS) results of 
comparable LCeMO films \cite{wer09}.
Furthermore, equation~(\ref{hopping_energy_vs_Mn_valence}) is only valid when we 
assume the equivalence of Mn$^{2+}$ and Mn$^{4+}$ with respect to the double 
exchange mechanism, as proposed in the theoretical work by Schlottmann \cite{Schlottmann2008}, which might also be questionable due to the 
high ionic radius of Mn$^{2+}$ and the subsequent dramatic decrease of the 
Mn-O-Mn bond angle. Indeed such a electron-hole asymmetry regarding the double exchange in LCeMO was proposed by Middey et al. \cite{Middey2014}. From these considerations, we have to state that both, the 
derivation of the Mn$^{3+}$ concentration from the polaron hopping energy as well 
as from the XPS data, have to be understood as estimates.

The \emph{comparative} analysis of the \textit{R-T} data within the temperature range between 300~K and 90~K, testing the APH and the 
VRH models, is depicted in figure~\ref{fig3} 
(exemplified for the 100-nm samples C and D).

For \emph{sample C}, as shown in figure~\ref{fig3}(a), from 210~K 
towards lower temperatures the resistance increase with decreasing 
temperature is slower than expected 
from the APH model. The Mott-VRH model according to 
eq.~(\ref{sigma_VRH_Mott}) is not 
appropriate as well, since plotting of $\ln(R/T^{1/4})$ as a function 
of $T^{-1/4}$ does not result in a linear dependence. 
However, the modified Efros-Shklovskii 
VRH model [eqs.~(\ref{sigma_mod_ES_VRH}), (\ref{T_0_mod_ES_VRH})] 
reproduces the measured data in the whole considered temperature range 
very well. Accordingly, plotting $\ln(R/T^{9/2})$ versus $T^{-1/2}$ 
indeed results in a linear dependence.
The fitting procedure yields a temperature independent $T_{0}=1.95\times 10^5$~K. 
According to equation eq.~(\ref{T_0_mod_ES_VRH}) this means that the polaronic gap 
$\Delta_{pg}$ is proportional to $T^{1/2}$ like reported in refs.~\cite{lai02, Laiho05}.
Since the APH model starts to deviate from the measured data at $T_V=$210~K it is appropriate 
to estimate $\Delta_{pg}$ as \cite{Laiho05}:
\begin{equation}
\label{polarongap}
\Delta_{pg} \approx E_h \left(\frac{T}{T_V}\right)^{1/2}  \quad .
\end{equation} 
With equation~(\ref{T_0_mod_ES_VRH}) and $E_h$ from table~\ref{tab_2} this yields $T_{ES}=1.22\times 10^5$~K.

Similarly, 
the transport data for \emph{sample B} was evaluated and we deduce 
 $T_{0}=1.92\times 10^5$~K and $T_{ES}=1.06 \times 10^5$~K (see also table~\ref{tab_2}). 
We stress that 
the Efros-Shklovskii VRH model is only valid for hopping energies 
$E_{ES}(T)$ [cf. eq.~(\ref{E_ES})] below the Coulomb gap $\Delta_C$ 
[eq.~(\ref{Delta_C})]. Since 
even at 300~K no significant deviations between experimental and 
theoretical data occur, the Coulomb gap can be estimated as 
$\Delta_C \approx E_{ES}^{300~K}=k_B \sqrt{T_{ES}\cdot 300~K}$, 
resulting in values of 0.41~eV and 0.44~eV for samples B and C, 
respectively, which is in good agreement with the values for 
La$_{0.8}$Ca$_{0.2}$MnO$_3$ reported in 
ref.~\cite{biswas_density_1999}. For an estimate of the electron 
localization length according to eq.~(\ref{T_ES}) we need the permittivity, 
which we assume to be in the range of $\epsilon_r=10\ldots 20$ according 
to literature values \cite{neupane_doping_2006} for La$_{1-x}$Ca$_x$MnO$_3$.
For samples B and C, we can now calculate localization lengths between 
0.3~nm and 0.6~nm. Thus, the electrons are delocalized over several unit 
cells, as expected for magnetic polarons \cite{coe95,vir97}.

For \emph{sample D}, the resistance as a function of the temperature at 
a measuring voltage of 1~V is depicted in fig.~\ref{fig3}(b). Between 
300~K and 260~K the data is rebuilt by the APH model with a hopping 
energy of 52~meV. For a measuring voltage of 5~V, the same 
fitting procedure between 300~K and 280~K gives a slightly larger hopping 
energy of 62~meV. According to eq.~(\ref{hopping_energy_vs_Mn_valence}) 
we obtain Mn$^{3+}$ concentrations 
of 68\% for $U=$~1~V and 70\% for $U=$~5~V. Assuming a true electron doping, 
i.e. a mixed Mn$^{2+/3+}$ valence, 
this would result in Mn valences of 2.68 or 2.70, which is -- in contrast 
to the cases of samples B and C -- lower than the XPS-derived 
value \cite{thiessen_mn2+/mn3+_2014} of 2.83. A possible explanation might be
that the high defect density in sample D supports a more uniform oxygen 
diffusion during the annealing process and not a reduction limited to the 
surface as for B and C.

For temperatures lower than 260~K the APH model fails to fit the 
experimental data, while, as shown in fig.~\ref{fig3}(b), the Mott-VRH 
model eq.~(\ref{sigma_VRH_Mott}) reproduces the data acquired at 1~V measuring voltage 
very well down to 90~K, giving a VRH temperature of 
$T_M=$~6.61$\times$10$^8$~K. For the data measured at 5~V the fitting procedure 
results in $T_M=$3.54~$\times$10$^8$~K, in the temperature range between 
280~K and 190~K. According to eq.~(\ref{E_M}) the average hopping energies 
read as 0.9~eV at 260~K and 0.37~eV at 90~K, which is decisively higher than 
the values for samples B and C, and must be explained with the stronger 
disorder in sample D (existence of microscopic CeO$_2$ clusters). Assuming a 
Coulomb gap of around 0.4~eV for sample D, too, it becomes clear why the 
ES-VRH model cannot reproduce the data appropriately: for that, the 
Coulomb gap would have to be larger than the hopping energy. The 
question why the activation energy jumps at 260~K, where we have changed 
the model for fitting the experimental data, from 52~meV (APH) to 900~meV 
(VRH) remains unsolved. One may speculate that the APH model is per se
unsuitable for such a strongly disordered manganite film.  

Finally, an estimate of the carrier localization length is made for 
sample D, too. For applying eq.~(\ref{T_M}) the density of states at the 
Fermi energy is needed, which we extract from the formula given by 
Viret et al. \cite{vir97} as follows:

\begin{equation}
\label{DOS}
N(E_F)=\frac{0.5 x (1-x)}{V_{uc} U_H} \quad ,
\end{equation} 
 
with $(1-x)$ being the concentration of the Mn$^{3+}$ ions, $V_{uc}$ 
being the unit cell volume, and $U_H=$2~eV the Hundt coupling energy. 
With $x=$0.17 from our former XPS measurements \cite{thiessen_mn2+/mn3+_2014}
and a lattice constant of $c=$0.3894~nm, a density of states 
of $N(E_F)=$5.95$\times$10$^{26}$~m$^{-3}$eV$^{-1}$ is calculated, 
which results in localization lengths of 0.09~nm and 0.11~nm for 
measuring voltages of 1~V and 5~V, respectively. At 260~K, the average 
hopping distance according to eq.~(\ref{E_M}) is 3.6~nm for both voltages, 
i.e., the electrons are mainly localized at the Mn ions and have to 
hop over approximately 10 lattice constants to find a place with 
suitable energy.
For an overview of the numerical data extracted from fitting the 
transport data of the highly reduced films refer to table~\ref{tab_2}.

\section{Summary}

The resistance-vs.-temperature characteristics of La$_{0.7}$Ce$_{0.3}$MnO$_3$ 
films of different thickness, oxygen reduction, and CeO$_2$ phase 
segregation have been recorded in the range between 300~K and 90~K. 
While the film thickness regarded here (up to 100~nm) has only a slight influence on the 
electrical transport, successive oxygen reduction leads -- as observed in other 
manganite compounds as well -- to a dramatic increase of the resistance and 
the loss of the manganite-typical metal-insulator transition.
As expected from a number of previous results, 
only the highly reduced LCeMO films exhibited a Mn valence 
below 3$+$ and thus an effective electron doping. Depending on the degrees 
of oxygen deficiency and CeO$_2$ segregation as well as on the temperature range, 
very individual states of disorder are prepared in the 
respective LCeMO films, which result in different transport mechanisms 
with widely scattered values for the hopping energies and the carrier localization 
lengths.

\section*{Acknowledgement}

This work was kindly supported by the German Science 
Foundation (DFG, grant no. BE 3804/2-1).

\section*{References}


\end{document}